\documentclass[12pt]{article}
\usepackage[margin=0.95 in]{geometry}
\usepackage{amsmath} 
\usepackage{amssymb,amsfonts}
\usepackage[all]{xy}
\usepackage{graphicx}
\usepackage{tikz-cd}
\usepackage{amsmath}
\usepackage{amssymb}
\usepackage{float}
\usepackage{array}
\usepackage{tikz}
\usepackage{mathtools}
\usepackage{mathrsfs} 
\usepackage{comment}
\usepackage[hidelinks]{hyperref}
\usepackage{cite}
\usepackage[multiple]{footmisc}
\usepackage{physics}
\numberwithin{equation}{section}
\begin{document}

\begin{titlepage}
\vbox{
    \halign{#\hfil         \cr
          } % end of \halign
      }  % end of \vbox
\vspace*{15mm}
\begin{center}
{\Large \bf 
More On  Pure Gravity with a }

 \vspace*{3mm}

 {\Large \bf  Negative Cosmological Constant
 }

\vspace*{15mm}

{\large  Lior Benizri and Jan Troost$^a$}
\vspace*{8mm}

$^a$ Laboratoire de Physique de l'\'Ecole Normale Sup\'erieure \\ 
 \hskip -.05cm
 CNRS, ENS, Universit\'e PSL,  Sorbonne Universit\'e, \\
 Universit\'e de Paris 
 \hskip -.05cm F-75005 Paris, France	 
\vskip 0.8cm
	% {\small
	% 	E-mails:
	% 	\texttt{lior.benizri@phys.ens.fr, jan.troost@ens.fr}
	% }
\vspace*{0.8cm}
\end{center}

\begin{abstract}
We identify an ambiguity in the Chern-Simons formulation of three-dimensional gravity with negative cosmological constant that originates in an outer automorphism of the Lie algebra $\mathfrak{sl}(2,\mathbb{R})$. It has important consequences for the stability of the theory in a space-time with boundary. We revisit the classical equivalence of three-dimensional gravity with a boundary Liouville theory both on and off the mass shell. Moreover, we provide further details on the quantum equivalence, the gauge symmetry that renders the spectrum diagonal, as well as the relation between asymptotically $AdS_3$ metrics and polar boundary conditions. We thus set the proposal that the Liouville conformal field theory  serves as a definition of a unitary theory of pure gravity in three dimensions with negative cosmological constant on a more stable footing. 
\end{abstract}
\end{titlepage}
\tableofcontents

\section{Introduction}
The  quantization of the theory of general relativity is largely an unsolved puzzle despite steady  progress over the last century. It may even be that the question is ill posed in the sense that we are attempting to quantize a thermodynamic master field. One traditional manner to nevertheless attack the problem is to reduce the number of degrees of freedom in the gravitational field even further by diminishing the dimension of space-time \cite{Deser:1983tn,Deser:1983nh}. When we move from four dimensions to three, we  render the theory of gravity topological \cite{Deser:1983tn,Deser:1983nh,Witten:1988hc}. The topological theory still has interesting observables. Moreover, it contains physical degrees of freedom when we study the theory on a manifold with boundary. 
Thus, it is worthwhile to quantize three-dimensional gravity, preferably in accordance with basic principles, among which unitarity. 

In this paper, we further investigate a concrete proposal to quantize pure gravity with
 a negative cosmological constant. Asymptotically $AdS_3$ space-times come equipped with a conformal boundary which  harbours degrees of freedom \cite{Brown:1986nw}. These degrees of freedom  interact as in a two-dimensional conformal field theory \cite{Maldacena:1997re} which we may hope to  use as a definition of pure gravity in three dimensions. 
Most attempts at defining such a theory of pure gravity assume that the global $AdS_3$ space-time is dual to a normalizable $\mathfrak{sl}(2,\mathbb{R}) \oplus \mathfrak{sl}(2,\mathbb{R})$  invariant ground state of the dual conformal field theory and that the spectrum of the dual conformal field theory is discrete -- see e.g. \cite{Witten:2007kt,Maloney:2007ud}.
These are natural assumptions, but
it is hard to make such theories consistent with fundamental properties of the  conformal field theory like invariance under large diffeomorphisms, its existence on general Riemann surfaces, and unitarity (see e.g. \cite{Gaberdiel:2007ve,Maloney:2007ud} as well as e.g. \cite{Raeymaekers:2020gtz}). 
In this work, we further study the very different possibility of using the unitary Liouville conformal field theory on the boundary of the space-time  \cite{Coussaert:1995zp} as a definition of a unitary theory of pure gravity with negative cosmological constant \cite{Li:2019mwb}. The properties of the resulting quantum theory of gravity are then entirely distinct \cite{Li:2019mwb}.

In order to pursue this possibility, we need to come to terms with various features of the  link between Liouville theory and Einstein gravity. The first feature is that the standard link leads to an unstable  boundary Liouville theory \cite{Coussaert:1995zp,Henneaux:1999ib}. 
A second aspect that we wish to master is the off-shell equivalence of the theories at the quantum level. To  these ends, we exploit the  literature on the subject and complement it where necessary. 

The plan of the paper is as follows. In section \ref{ThreeDimensionalGravity} we briefly review the Chern-Simons formulation of three-dimensional gravity. We point out an important ambiguity in its definition due to the existence of an outer automorphism of the Lie algebra $\mathfrak{sl}(2,\mathbb{R})$. In section \ref{ClassicalLiouvilleGravity} we review the classical equivalence between various formulations of the Chern-Simons theory of gravity with negative cosmological constant. We  clarify the gauge symmetries of the model on a manifold with boundary and the role of the outer automorphism in determining the 
conformal field theory on the boundary, and its stability. We moreover provide a detailed map between on the one hand, the space of classical solutions which gives rise to the Hilbert space of the quantum Liouville theory after quantization and on the other hand, a definite  subset of metric solutions to Einstein gravity. 
Finally, in section \ref{QuantumLiouvilleGravity} we exploit the preparatory work to provide more details on the quantum equivalence of a unitary theory of gravity with negative cosmological constant in three dimensions and the Liouville conformal field theory.  We conclude in section \ref{Conclusions} with a recap of the distinct features of this quantum theory of gravity and indications on how it might be put to use to analyze features of quantum gravity in general, or of this  theory of quantum gravity in particular. 

\section{Three-dimensional Gravity with a Twist}
\label{ThreeDimensionalGravity}
\label{Outer}
We briefly recall  the Chern-Simons formulation of three-dimensional gravity \cite{Achucarro:1986uwr,Witten:1988hc}. We point out an important ambiguity in the map between the dreibein and the spin connection on the one hand and the Chern-Simons fields on the other, rooted in an outer automorphism of the  Lie algebra $\mathfrak{sl}(2,\mathbb{R})$. 

\subsection{The Metric and the Chern-Simons Theories}
The metric $g$ is related to a dreibein one-form $e=e^a t_a$ through:
\begin{equation}
g = 2 \Tr (e \otimes e) \, ,
\end{equation}
where the trace is normalized such that $ 2 \Tr (t_a t_b)=\eta_{ab}$. 
 In three dimensions, the spin connection $\omega$ can be dualized to a $\mathfrak{sl}(2,\mathbb{R})$ Lie algebra valued one-form $\omega=\omega^a t_a$.
The action $S_{CS}$ of Chern-Simons theory with gauge field $A$ is:
\begin{equation}
	S_{CS}[A]= \frac{k}{4 \pi} \int d^3 x \Tr \left(A\wedge dA +\frac{2}{3}A\wedge A \wedge A\right) \, .
\end{equation}
 The gauge
algebra considered in the Chern-Simons formulation of three-dimensional gravity with a negative cosmological constant is $\mathfrak{sl}(2,\mathbb{R})\oplus \mathfrak{sl}(2,\mathbb{R})$. The Riemann-Hilbert action can be
identified with the difference of two Chern-Simons actions, each with gauge algebra $\mathfrak{sl}(2,\mathbb{R})$ \cite{Achucarro:1986uwr,Witten:1988hc}:
\begin{equation}
	S_{CS}[A_L,A_R]=S_{CS}[A_L]-S_{CS}[A_R] \, ,
\end{equation}
where $k$ is classically related to the radius of curvature $l$ and the Newton constant $G$ through the relation $k=l/(4G)$. Before we define the gauge fields $A_{L,R}$ in terms of the dreibein $e$ and spin connection $\omega$, we need to discuss an aspect of the gauge algebra that gives rise to an ambiguity in the identification. 
\subsection{An Outer Automorphism}
\label{OuterAutomorphism}
The $\mathfrak{sl}(2,\mathbb{R})$ Lie algebra has automorphism group $PGL(2,\mathbb{R})$. The group $PGL(2,\mathbb{R})$ has two components. The identity component coincides with $PSL(2,\mathbb{R})$. The elements with negative determinant make up a second component. In the  two-dimensional, fundamental representation of the Lie algebra $\mathfrak{sl}(2,\mathbb{R})$ which we specify in Appendix \ref{Conventions}, we can define an element $\alpha$ of the second component through conjugation of the Lie algebra elements by the $PGL(2,\mathbb{R})$ matrix
\begin{equation}
h_\alpha = \begin{pmatrix}
	-1&0\\
	0&1
	\end{pmatrix} \, .
 \end{equation}
 The $PGL(2,\mathbb{R})$ group element $h_\alpha$ has determinant minus one.  
For a $\mathfrak{sl}(2,\mathbb{R})$ Chern-Simons theory, we can consider various gauge groups, like $G=PSL(2,\mathbb{R})=SO(2,1)$, $SL(2,\mathbb{R})$, or its universal cover $\widetilde{SL(2,\mathbb{R})}$. In all these cases,  the outer automorphism $\alpha$ generates a global $\mathbb{Z}_2$ symmetry of the theory. 

\subsection{The Chern-Simons Formulations of Gravity}
The traditional identification between the Chern-Simons fields $A_{L,R}$ and the dreibein $e$ and spin connection $\omega$ is: 
\begin{align}
A_L &= \omega + \frac{e}{l}
\nonumber \\
A_R &= \omega - \frac{e}{l} \, . \label{ConnectionIdentity}
\end{align}
It is important for our purposes that another choice is equally natural. We can identify:
\begin{align}
A_L &= \omega + \frac{e}{l}
\nonumber \\
\alpha(A_R) &= \omega -\frac{e}{l} \, , \label{ConnectionOuter}
\end{align}
where $\alpha(A_R)$ equals the connection $A_R$ after the action of the outer automorphism $\alpha$, i.e. $\alpha(A_R)=h_\alpha A_R h_\alpha^{-1}$.  
The metric then either takes the form:\footnote{In the following we often set $l=1$ for convenience.}
\begin{equation}
g_{\text{Id}} = \frac{1}{2} \Tr ( (A_L-A_R) \otimes (A_L-A_R))
\,  \label{MetricIdentity}
\end{equation}
or 
\begin{equation}
g_{\alpha} = \frac{1}{2} \Tr  ( (A_L-\alpha (A_R)) \otimes (A_L-\alpha (A_R))) \, .
\label{MetricOuter}
\, 
\end{equation}
These expressions differ. 
While $\alpha$ is a global symmetry of either of the Chern-Simons theories separately, it plays an important role when we define non-chiral quantities. The metric $g$ is an  example of such a non-chiral object. 
In the analysis that follows, we will mostly study the left and right chiral theories  separately. However, we will always keep in mind the possibility to twist one of the two chiral theories by an outer automorphism. This will have  important  consequences for our interpretation of the combination of the chiral theories as a theory of gravity.

\section{Classical Three-dimensional Liouville Gravity}
\label{ClassicalLiouvilleGravity}
General relativity with negative cosmological constant and $AdS_3$ boundary conditions has been related to Liouville theory on the boundary \cite{Coussaert:1995zp}. We review this relation in this section. We emphasize the  implementation of the boundary conditions by integrating over  Lagrange multiplier gauge fields  and how these give rise to gauge symmetries on the boundary. We identify the boundary conditions corresponding to a conformal class of boundary metrics and remark on their similarity to opers.
We moreover identify the subset of classical solutions which are quantized in   Liouville theory.
The whole of the analysis is geared towards clarifying the equivalence of a quantum Liouville theory and a three-dimensional theory of gravity in the bulk.  

\begin{figure}
\begin{center}
\begin{tikzcd}[column sep=50pt]
\text{Liouville Theory} \arrow[<->]{d} \\
\text{Null Gauged Wess-Zumino-Witten} \arrow[<->]{d} \\
\text{Chiral and Anti-chiral Null Gauged Wess-Zumino-Witten} \arrow[<->]{d} \\
\text{Two Chern-Simons Theories with Extended Boundary Conditions} \arrow[<->]{d} \\
\text{Einstein-Hilbert Theory in Asymptotically } AdS_3
\end{tikzcd}
\end{center}
\caption{The path from boundary to bulk and back in Liouville pure three-dimensional gravity with negative cosmological constant.}
\label{Path}
\end{figure}
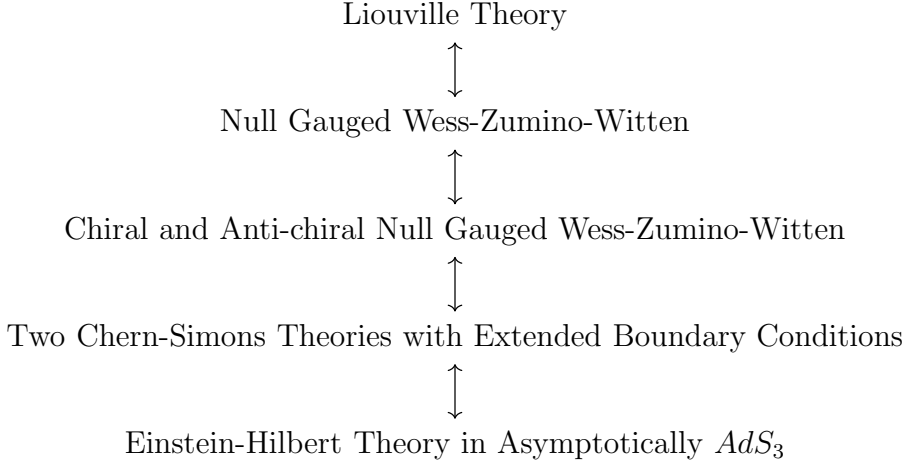

Since this section contains a large number of stepping stones that relate the boundary Liouville theory to the bulk Einstein-Hilbert theory of gravity \cite{Coussaert:1995zp}, it may be useful to have the schematic plan of the list of theories in mind. We cycle through the equivalences from top to bottom in Figure \ref{Path}.
At each step we  keep track of the gauge symmetries that play a crucial role. We also carry along a  large set of classical solutions  and list them in particular gauges for each theory. We wind up with a particular  subset of  asymptotically $AdS_3$ metric solutions to Einstein-Hilbert gravity that form the classical phase space of our quantum theory of gravity  \cite{Li:2019mwb}.

\subsection{Liouville Theory}
We start out by considering a Liouville conformal field theory on the two-dimensional plane \cite{Liouville,Seiberg:1990eb}.
It is a unitary theory with consistent spectrum and three-point functions that determine the full theory on any Riemann surface \cite{Dorn:1994xn,Zamolodchikov:1995aa,Teschner:2001rv}.\footnote{Moreover, it is important to realize that it is hard to modify either the spectrum or the three-point functions in a manner respecting conformal field theory axioms. See also the bootstrap \cite{Collier:2017shs} for  arguments in favour of the uniqueness of the theory. The analytic uniqueness arguments are strongest for values of $b^2$ which are irrational.}       
 The action of  the  Liouville scalar field $\phi$ is:\footnote{Our light-cone coordinates are defined through through the relations $x^\pm = t \pm x$.}
\begin{equation}
	S_{\text{Liouville}} =\frac{1}{2 \pi}
 \int d^2  x\, \left(\frac{1}{2}\, \partial_+ \phi\, \partial_- \phi - M \exp \sqrt{2} b  \phi \right) \, .
\end{equation}
We assume that there is a linear dilaton  coupling to the two-dimensional curvature tensor which gives the theory a central charge $c=1+6 Q^2$ where $Q=b+b^{-1}$. The exponential potential is marginal in the quantum theory.
The constant $M$ can be absorbed through a shift of the Liouville field $\phi$ such that only its sign is important.\footnote{We consider the cases $M \neq 0$.}
The equation of motion for the scalar field $\phi$ is the Liouville equation:
\begin{equation}
	\partial_+ {\partial}_- \phi+\sqrt{2} b Me^{\sqrt{2} b \phi}=0 \, .
 \label{LiouvilleEOM}
\end{equation}
\subsubsection*{The Classical Liouville Solutions}
The general solution to the Liouville equation can be parameterized in terms of two monotonic functions $F(x^+)$ and  $\tilde{F}(x^-)$ of the light-cone coordinates $x^\pm$ \cite{Liouville}: 
\begin{equation}
	e^{\sqrt{2} b \phi}=\frac{1}{b^2 M}\frac{F' \tilde{F}'}{(1+F \tilde{F})^2} \, ,
 \label{LiouvilleSolutions}
\end{equation}
where the prime denotes a derivative with respect to the argument of the functions. The 
 two functions $F$ and $\tilde{F}$ must have a product of derivatives which has the sign of the parameter $M$ in the action since the field $\phi$ is real and its exponential is positive. This is the large set of classical solutions that we will track along our road through various reformulations of the classical theory.\footnote{It is a larger set than the classical phase space we quantize -- we will restrict to the appropriate subspace at the end of this section. }
\subsection{The Null Gauged Wess-Zumino-Witten Model}
The Liouville conformal field 
theory is equivalent to a null-gauged Wess-Zumino-Witten model \cite{Forgacs:1989ac,Balog:1990mu}. The Lagrangian of the gauged model is \cite{Balog:1990mu}:
\begin{align}
S_{GWZW}[g,C_L,C_R]=S_{WZW}[g]+\frac{k}{ \pi}\int & d^2 x  \{C_L
 \left( (\partial_+ g\, g^{-1})^- -\mu \right) \nonumber \\
	&-C_R 
 \left( (g^{-1} {\partial}_- g)^+ -\nu  \right)- \Tr (C_L t_+ g C_R t_- g^{-1}) \} \, ,
 \label{GaugedWZWAction}
\end{align}
where $g$ is a $SL(2,\mathbb{R})$ group-valued field and $C_{L,R}$ are gauge fields in the null directions  $t_{\pm}$ inside the algebra $\mathfrak{sl}(2,\mathbb{R})
%\oplus \mathfrak{sl}(2,\mathbb{R})
$.\footnote{See Appendix \ref{Conventions} for our conventions for the Lie algebra generators.}
%%% 
We have used the $t_-$ component of the current $\partial_+ g g^{-1} = (\partial_+ g g^{-1})^a t_a$ in the action and the constant $\mu$ can also be thought off as lying in the $t_-$ direction of the Lie algebra. 
The gauge field ${C}_L$ couples to the current that generates the left chiral symmetry while the gauge field $C_R$ couples to the right algebra. 
The quantum central charge of the model with improved energy momentum tensor (and ghosts for gauge fixing) is $c=3k/(k-2)-2 +6 k =1+6 (\sqrt{k-2} + 1/\sqrt{k-2})^2$ -- see e.g. \cite{Bershadsky:1989mf}. 
We can pick the relation $b^{-1}=\sqrt{k-2}$ (which is one out of four choices) between the Liouville parameter $b$ and the level of the Wess-Zumino-Witten model. 
  The action $S_{WZW}[g]$ is the standard Wess-Zumino-Witten action: 
  % 
  % Summary: with coefficient k/4pi and the trace in the fundamental representation, the Chern-Simons action is properly quantized for su(n) when k is an integer. 
  %
\begin{equation}
S_{WZW}[g]=\frac{k}{2 \pi}\int d^2 x\ \Tr[g^{-1} \partial_+ g  g^{-1} \partial_- g]+\frac{k}{12 \pi}\int d^3 x\, \Tr[ (\ast \hat{g}^{-1}d\hat{g})^3]
\end{equation}
where $\hat{g}$ is an extension of $g$ over a three-dimensional space-time with boundary equal to our two-dimensional space-time and we pull back the one-form through the group valued map. We will use the notation $S_{WZ}(\hat{g})$ for the second (Wess-Zumino) term.
The  action is gauge invariant under the  transformations \cite{Balog:1990mu}:
\begin{align}
	 g & \to \alpha g \beta^{-1} 
 \nonumber 
 \\
	C_{L} & \to \alpha C_{L}\alpha^{-1}+\alpha\partial_{-}\alpha^{-1} 
 \nonumber
 \\
	 C_{R} & \to \beta C_{R}\beta^{-1}+ \beta \partial_+ \beta^{-1}
\end{align}
where we have local group valued gauge parameters $\alpha = e^{\alpha^{+}t_{+}}$ and $\beta =e^{\beta^- t_{-}}$ .

To connect this theory to Liouville theory, we Gauss decompose the group $SL(2,\mathbb{R})$. Any element $g$ of $SL(2,\mathbb{R})$ can locally 
be written as:
\begin{equation}
	g=\exp(xt_{+})\exp\left(
 \sqrt{2} b\phi\, t_2\right)\exp(yt_{-}) \, .
\end{equation}
The left gauge transformation thus translates the group coordinate $x$ at will, while the right multiplication translates the field $y$. Only the field $\phi$ is physical. To see the equivalence to the Liouville theory, we choose the  
gauge $C_{L,R}=0$ in which the equations of motion read:
\begin{align}
	\partial_+ (g^{-1}{\partial}_- g) &=0 \label{WZWEOM}\\
	%&J \coloneqq 
  \, \Tr (t_{+} \partial_+ g\, g^{-1} ) &= \mu \label{J1}\\
%	&\bar{J} \coloneqq 
  \, \Tr (t_{-} g^{-1}{\partial}_- g ) &= \nu\label{J2} \, .
\end{align}
In coordinates, after eliminating the fields $x,y$ through their constraints, we find the equation of motion for the Liouville field 
(\ref{LiouvilleEOM}) under the identification:
\begin{equation}
b^2 M = \mu \nu \, .
\end{equation} This shows the classical equivalence of the models.
\subsubsection*{The Null Gauged Classical Solutions}
The classical solutions to Liouville theory take on a simple form in the gauged Wess-Zumino-Witten model and can be  derived simply in the latter \cite{Forgacs:1989ac}. 
The most general solution of the model in the $C_{L,R}=0$ gauge is:
\begin{equation}
	g(x^+,x^-)=g_+(x^+) \, g_-(x^-)  \, , \label{WZWsol}
\end{equation}
where $g_+$ and $g_-$ are chiral $SL(2,\mathbb{R})$-valued functions satisfying the constraints  (\ref{J1}) and
(\ref{J2}). 
After Gauss decomposition of the left and right group valued functions,  the constraints (\ref{J1}) and (\ref{J2})  read:
\begin{equation}
	 \partial_+ y_{+} =\mu e^{\sqrt{2} b\phi_+} \, , \qquad
	 {\partial}_- x_{-} =\nu e^{\sqrt{2} b \phi_-} \, .
\end{equation}
We can solve for the on-shell Liouville chiral fields $\phi_{\pm}$:
\begin{equation}
	\sqrt{2} b \phi_{+}  =\log \left( y'_{+}/ \mu \right)\, , \qquad 
	\sqrt{2} b \phi_{-}  =\log \left( x'_{-}/ \nu \right) \, .
\end{equation}
The general solution $g$ to the gauged Wess-Zumino-Witten model is thus fully parameterized by two
chiral functions $y_{+}$ and $x_{-}$. The $\phi$ field in the Gauss decomposition of the group valued function $g$
is identified with the Liouville field and reads:
\begin{align}
	e^{\sqrt{2} b \phi}=\frac{1}{b^2 M}\frac{y'_{+}x'_{-}}{(1+y_{+}x_{-})^2} \, .
\end{align}
The solution to the gauged model depends on a choice of gauge. It will be convenient for us to work in a gauge where the following equations hold:
\begin{equation}
	x_+=-\frac{y_{+}''}{2\mu y_{+}'}, \, y_-=-\frac{x_{-}''}{2\nu x_{-}'} \, .
	\label{GFcondition}
\end{equation}
From the Wess-Zumino-Witten solution (\ref{WZWsol})  this gauge is reached through  the  transformation with gauge parameters:
\begin{equation}
	\alpha=\exp\left[-\left(x_+ +\frac{y_+''}{2\mu y_+'}\right)t_+\right], \, \beta=\exp\left[-\left(y_{-}+\frac{x_-''}{2\nu x_-'}\right)t_-\right] \, .
\end{equation}
The constraints,
the currents and the choice of gauge $C_{L,R}=0$ are invariant under these residual chiral gauge transformations.
After fixing the  gauge in the manner described, we can extract a solution to the Wess-Zumino-Witten model from a Liouville field through the 
identification $y_+\equiv F$ and $x_{-}\equiv \tilde{F}$:
\begin{equation}
	g=\sqrt{\frac{\mu \nu}{F'\tilde{F}'}}\begin{pmatrix}
		\dfrac{2(F')^2-F F''}{2\mu F'}	&-\dfrac{F''}{2\mu F'}\\
		F&1
	\end{pmatrix}
	\begin{pmatrix}
		\dfrac{2(\tilde{F}')^2-\tilde{F} \tilde{F}''}{2\nu \tilde{F}'}& \tilde{F}\\
		-\dfrac{\tilde{F}''}{2\nu \tilde{F}'}&1
	\end{pmatrix} \, .
\label{OnShellGroupElement}
\end{equation}
Thus, we proved the classical equivalence along the first step in Figure \ref{Path} and we tracked a large set of classical solutions.

\subsection{Chiral Gauged Wess-Zumino-Witten Models}
The next step we take towards the chiral Chern-Simons actions is to divide the non-chiral Wess-Zumino-Witten model (\ref{GaugedWZWAction}) into a purely left- and a purely right-moving gauged Wess-Zumino-Witten
model. This intricate process is reviewed in Appendix \ref{WZWSplit}. See also \cite{Stone:1989vg,Harada:1990tw}. 
The upshot is that we have a chiral and an anti-chiral gauged model with fields $(g_L,g_R)$ related to the field $g$ through the identification
\begin{equation}
    g=g_L g_R \, ,
\end{equation}
and with chiral actions: 
\begin{align}
S_L [g_L,C_L]&= \frac{k}{2 \pi} \int d^2 x \Tr [ 
g_L^{-1} \partial_- g_L g_L^{-1} \partial_x g_L
]+S_{WZ}(\hat{g}_L) \nonumber \\
&
 + \frac{k}{\pi} \int d^2 x \Tr[ C_L ( (\partial_x g_L g_L^{-1})^- - \mu)  ]
\label{ChiralGaugedWZW}
\end{align}
and 
\begin{align}
S_R[g_R,C_R] &= -\frac{k}{2 \pi} \int d^2 x \Tr [ g_R^{-1} \partial_+ g_R g_R^{-1} \partial_x g_R  ] +S_{WZ}(\hat{g}_R) \nonumber \\
& + \frac{k}{\pi} \int d^2 x \Tr[ C_R ((g_R^{-1} \partial_x g_R)^+ + \nu)] \nonumber
\, .
\end{align}
The split of the Wess-Zumino-Witten field leaves its imprint on the chiral fields through the local multiplication symmetry
\begin{align}
	g_L &\to  g_L k^{-1} \nonumber \\
	g_R &\to k g_R \, , \label{LeftMultiplication}
\end{align} 
where $k(x^+,x^-)$ is an arbitrary group valued function.
Indeed, the split has a large degree of arbitrariness. One way to summarize these facts is to say that the action $S_{GWZW}$ describes the same theory as the action $S_L+S_R$ with the left-multiplication symmetry (\ref{LeftMultiplication}) gauged  \cite{Stone:1989vg,Harada:1990tw}. We therefore postulate throughout that the latter symmetry is indeed gauged.

\subsubsection*{Chiral Gauged Classical Solutions}
On-shell, there
is a canonical split of the left and right chirality group elements, modulo a constant group element. It is already exhibited in the solution for the group valued map (\ref{OnShellGroupElement}). Applying the split to our set of classical solutions, we find  solutions to the chiral gauged Wess-Zumino-Witten models: 
\begin{equation}
	g_L=\sqrt{\frac{\mu}{F'}}
    \renewcommand{\arraystretch}{1.5}
    \begin{pmatrix}
		\dfrac{2(F')^2-F F''}{2\mu F'}	&-\dfrac{F''}{2\mu F'}\\
		F&1
    \end{pmatrix},\, 
	g_R=\sqrt{\frac{\nu}{\tilde{F}'}}
    \renewcommand{\arraystretch}{2.2}
	\begin{pmatrix}
		\dfrac{2(\tilde{F}')^2-\tilde{F} \tilde{F}''}{2\nu \tilde{F}'}& \tilde{F}\\
		-\dfrac{\tilde{F}''}{2\nu \tilde{F}'}&1
	\end{pmatrix} \, .
\end{equation}
We have ignored possible subtleties with the square root -- if necessary one adds a sign in both the left-moving and right-moving square roots. 
Once more it is possible to check directly that these configurations solve the equations of motion of the chiral models.

\subsection{Two Chern-Simons Theories}
The third step in Figure \ref{Path} -- see also \cite{Coussaert:1995zp} -- is to exploit the relation between chiral Wess-Zumino-Witten models on a two-dimensional boundary and Chern-Simons theories on a three-manifold \cite{Witten:1988hf,Elitzur:1989nr}. To that end, we introduce a radial coordinate $r$ that moves inside a three-dimensional space-time and that goes to infinity as we approach the boundary. We wish to extend our two-dimensional group valued fields $g_L$ and $g_R$ into the bulk.  
The chiral Wess-Zumino-Witten actions can be rewritten as the difference of  two Chern-Simons actions, each with gauge algebra $\mathfrak{sl}(2,\mathbb{R})$ on a three-manifold $M$ whose boundary $\partial M$ is the two-dimensional space-time $\Sigma$ we started out with.
To prove the equivalence, and to connect to the gravity theory later on, we assume particular boundary conditions on the left and right fields. On the left, one boundary condition will be 
\begin{equation}
A_{L,-} =_{|_\partial} 0 \label{LeftParallelBC}
\end{equation}
and on the right
\begin{equation}
A_{R,+} =_{|_\partial} 0 \, . \label{RightParallelBC}
\end{equation}
These boundary conditions which set to zero a component of the gauge field along the boundary guarantee that the variational principle for the Chern-Simons theory on a manifold with boundary is well-defined \cite{Elitzur:1989nr}.
Given these boundary conditions, for the left-moving action, it is convenient to define the new coordinates:
\begin{equation}
(t',x',r')=(t,x+t,r) \, .
\end{equation}
The boundary condition (\ref{LeftParallelBC}) is then transformed into the more traditional boundary condition $A'_{t'}=0$ which allows us to directly recuperate the  result for the Chern-Simons action on a manifold with boundary written in terms of a chiral Wess-Zumino-Witten model \cite{Elitzur:1989nr}:
\begin{equation}
S_{CS}[A_L] = S_{WZ}(\hat{g}_L)+\frac{k}{4 \pi} \int_{\partial M} Tr (g_L^{-1} \partial_{x'} g_L g_L^{-1} \partial_{t'} g_L) \, , 
\end{equation}
where the other gauge field components 
\begin{equation}
\tilde{A}_L=-\tilde{d} \hat{g}_L \hat{g}_L^{-1}
\label{SolutionConstraint}
\end{equation}
solve the Gauss constraint and the quantities with a tilde refer to the $x'$ and $r'$ directions only. 
The map $\hat{g}_L$ is a three-dimensional extension of the boundary map $g_L$. In the original coordinates, we then  have:
\begin{equation}
S_{CS}[A_L] =  S_{WZ}(\hat{g}_L)+\frac{k}{2 \pi} \int_{\partial M} Tr (g_L^{-1} \partial_{x} g_L g_L^{-1} \partial_{-} g_L) =S_L[g_L] \, ,
\end{equation}
where the last action is the chiral Wess-Zumino-Witten model action. 

The boundary term in (\ref{ChiralGaugedWZW}) proportional to the gauge field $C_L$ comes along for the ride.
The result is independent of the extension $\hat{g}_L$ of the group valued field $g_L$ to the three-dimensional bulk. 
For the right action, we set:
\begin{equation}
\tilde{A}_R = \hat{g}_R^{-1} \tilde{d} \hat{g}_R \, .
\end{equation}
This gives the relation:
\begin{equation}
S_{CS}[A_R] = -S_R[g_R]
\end{equation}
between the Chern-Simons and the anti-chiral Wess-Zumino-Witten action. In short, 
we find the equality: 
\begin{equation}
S_L[g_L,C_L]+S_R[g_R,C_R] = S_{CS}[A_L]-S_{CS}[A_R]+\frac{k}{\pi}  \int_{\partial M} \hspace{-3mm}d^2 x\; (-) C_L \left( A_{L,x}^-
%( 
%\partial_x g_L g_L^{-1})^- 
+\mu  \right) +
 C_R \left(
 %( g_R^{-1}\partial_{x}g_R)^+
 A_{R,x}^+
 +\nu \right) \, . \nonumber
\end{equation}
Let us briefly comment on how global and gauge symmetries interplay in the various formulations of our theory.  The gauge symmetries of the Chern-Simons theory became global symmetries on the boundary. The null algebras that we gauge on the boundary are embedded in these global symmetry algebras. 
On the other hand, the boundary gauge symmetry we use to glue the chiral and anti-chiral Wess-Zumino-Witten models is a hidden gauge symmetry of the combination of Chern-Simons theories on a manifold with boundary, restricted to an appropriate set of observables.

\subsubsection*{The Asymptotics}

The next stage is a more subtle and important step. Firstly, we slightly move our boundary into the bulk towards $r^{-1}=\epsilon>0$ where $\epsilon$ can later be interpreted as a bulk infrared cut-off. Secondly, we perform formal radial gauge transformations
\begin{equation}
A_L \rightarrow h^{-1}_L A_L h_L + h_L^{-1} d h_L
\end{equation}
\begin{equation}
A_R \rightarrow h^{-1}_R A_R h_R + h_R^{-1} d h_R
\end{equation}
with gauge parameters
\begin{equation}
h_L =\begin{pmatrix}
		\sqrt{r}&0\\
		0& \frac{1}{\sqrt{r}}
	\end{pmatrix}, \, \qquad
h_R=\begin{pmatrix}
		\frac{1}{\sqrt{r}}&0\\
		0& \sqrt{r}
	\end{pmatrix}  
\end{equation}
which are everywhere well-defined on the manifold with cut-off.
This has the effect of redefining the group valued radial extension:
\begin{equation}
	G_{L}=\begin{pmatrix}
		\frac{1}{\sqrt{r}}&0\\
		0& \sqrt{r}
	\end{pmatrix} \hat g_{L}, \, \qquad
	G_{R}= \hat g_{R}\, 
	\begin{pmatrix}
		\frac{1}{\sqrt{r}}&0\\
		0& \sqrt{r}
	\end{pmatrix} \, .
 \label{RadialGaugeTransformation}
\end{equation}
The  gauge transformation leaves the boundary condition $A'_{t'}=0$ invariant and therefore also our previous reasoning that links the Chern-Simons and the chiral Wess-Zumino-Witten actions. The gauge transformation induces a boundary term at the infinite past and future that influences the description of the asymptotic states, but it will not influence the theory on the radial boundary of the infinite cylinder. The formal gauge transformation does allow us to reproduce the asymptotics of $AdS_3$, in particular in the Fefferman-Graham gauge \cite{FG}. 
Indeed, originally, we start out with an asymptotic behaviour near the boundary at $r=\infty$ of the form 
\begin{equation}
A_{L,r} = 0 
  %\begin{pmatrix}
	% 	0&0\\
	% 	0 & 0
	% \end{pmatrix} 
 \, , \quad
 A_{L,-} = \begin{pmatrix}
		0&0\\
		0& 0
	\end{pmatrix} \, ,
\quad
A_{L,+} = \begin{pmatrix}
		0&O(1) \\
		-\mu& 0
	\end{pmatrix} \, .
\end{equation}
One can check that these boundary conditions are consistent with the equations of motion which require that the Chern-Simons connections be flat and do correctly implement the boundary condition, the constraint and the fixing of the bulk and boundary gauge symmetries. %
If we then perform the radius dependent gauge transformation (\ref{RadialGaugeTransformation}), we create a radial asymptotics $A_{L, r}=t_2/(2r)$ and $A_{R, r}=-
t_2/(2r)$ at large $r$.
In this gauge for $A_{L,r}$, the boundary conditions and radial behaviour can be summarized as:
\begin{equation}
	A_L\sim 
	\begin{pmatrix}
		\dfrac{dr}{2r}& \mathcal{O}(1/r)\\
		-\mu r dx^{+}& -\dfrac{dr}{2r}
	\end{pmatrix}
	,\, \qquad
	A_R \sim 
	\begin{pmatrix}
		-\dfrac{dr}{2r}&\nu r dx^{-} \\
		\mathcal{O}(1/r)& \dfrac{dr}{2r}
	\end{pmatrix} \, .
 \label{asymptotics}
\end{equation}
This matches the  $AdS_3$ boundary conditions \cite{Coussaert:1995zp}. We add the observation that the right upper entry of $A_L$ is proportional to $dx^+$ only, to order $O(1/r)$. This is true in the Chern-Simons formulation of the theory because the variational principle for the Chern-Simons action must be well-defined in the presence of a boundary. 
By the transformation of the components of the boundary gauge fields, the actions become: 
%Signs: if written in terms of $A_{R,x}$, put a +$\nu$.
\begin{align}
	S_{L}[A_L] &=S_{CS}[A_L]-\frac{k}{ \pi} \int d^2 x [C_L (A_{L,+}^-+\mu \, r)]\\
	S_{R}[A_R] &=S_{CS}[A_R]+\frac{k}{ \pi}\int d^2 x [C_R (A_{R,-}^+ -\nu \, r)]
\, . \end{align}
Finally, we note that these boundary conditions are a close cousin to an oper. Indeed, the boundary conditions: 
\begin{equation}
A_{op,r} = 0 
  %\begin{pmatrix}
	% 	0&0\\
	% 	0 & 0
	% \end{pmatrix} 
 \, , \quad
 A_{op,-} = \begin{pmatrix}
		0&0 \\
		0& 0
	\end{pmatrix} \, ,
\quad
A_{op,+} = \begin{pmatrix}
		0&0 \\
		-\mu& 0
	\end{pmatrix}
\end{equation}
were gauge transformed precisely in the manner above and discussed in \cite{Gaiotto:2011nm} in gauge invariant terms. We note that the double pole in asymptotically $AdS$ metrics  \cite{LeBrun:1982vjh}, which are directly related to the fact that the boundary carries a conformal class of metrics, is translated in the ('square root') Chern-Simons formulation as a simple pole boundary behaviour.  
The polar boundary conditions above may thus also allow for an invariant formulation of AdS boundary conditions in the Chern-Simons formulation of three-dimensional gravity. 

The equations (\ref{asymptotics}) define the set of allowed configurations in the gravitational theory. A crucial point is that all the gravitational configurations in our phase space are gauge inequivalent with respect to all the gauge symmetries discussed in the previous section.    
\subsubsection*{The Chern-Simons Classical Solutions}
Finally, when we trace the classical solutions to the bulk Chern-Simons theory after the radial gauge transformation,  we end up with the flat connections:
\begin{equation}
	A_L=
	\begin{pmatrix}
		\dfrac{dr}{2r}&\frac{\{F,x^+\}}{2\mu r}\,dx^+\\
		-\mu r\, dx^+& -\dfrac{dr}{2r}
	\end{pmatrix},
	\, \quad
	A_R=	
	\begin{pmatrix}
		\hspace{-3.5mm} -\dfrac{dr}{2r} & \nu r\, dx^-\\
		-\frac{\{\tilde{F},x^-\}}{2\nu r}dx^-  & \dfrac{dr}{2r}
	\end{pmatrix} \, .
 \label{ChernSimonsSolutions}
\end{equation}
In these matrices, the curly brackets $\{\,,\}$ stand for the
Schwartzian derivative:
\begin{align}
	\{F(x^+) , x^+ \} &=\frac{2 F' F'''-3(F'')^2}{2(F')^2}
 %\\
	%\{x^-,\tilde{F}\} %&=\frac{2 \tilde{F}' \tilde{F}'''-3(\tilde{F}'')^2}{2(\tilde{F}')^2}
 \, .
\end{align}
%with $F' \coloneqq \partial F,\, \tilde{F}'\coloneqq \bar{\partial} \tilde{F}$.
Note that we have taken $A_{L,-}$ and $A_{R,+}$ to vanish in the above. Given the other components of the gauge fields, this is necessary to satisfy the
 equations of motion of the  Chern-Simons theories.
We turn to the  final step in Figure \ref{Path} which is to translate these Chern-Simons fields into the metric theory of gravity. 

\subsection{The Einstein Hilbert Theory}
The Einstein-Hilbert theory of pure gravity has a classical equation of motion that demands that the metric in vacuum be Ricci flat. Moreover, one assumes that the connection has zero torsion. Under the field re-definitions (\ref{ConnectionIdentity}) or (\ref{ConnectionOuter}) between the dreibein $e$, spin-connection $\omega$ and the gauge fields $A_{L,R}$, these conditions are equivalent to the zero curvature equations of motion of the Chern-Simons theory. Thus, the classical theories are once more equivalent \cite{Achucarro:1986uwr,Witten:1988hc}. Moreover, the boundary conditions on the Chern-Simons theories lead to a prescribed asymptotic behaviour for the boundary metric \cite{Coussaert:1995zp}. 
We stressed in section \ref{Outer}  that in the identification between Chern-Simons gauge fields and the bulk metric, an outer automorphism plays an important role. It is finally time to prove this at the hand of the classical solutions and to discuss how it even affects the stability of the theory.
\subsubsection*{The Classical Solutions of Gravity with a Negative Cosmological Constant}
We can identify the classical solutions of gravity that correspond to the solutions (\ref{LiouvilleSolutions}) of Liouville theory.
We compute the metrics (\ref{MetricIdentity}) and (\ref{MetricOuter})  by plugging in the on-shell  Chern-Simons fields (\ref{ChernSimonsSolutions}): 
%\begin{equation}
%	g_{\mu\nu}=\eta_{ab} (A_{L,\mu}^{a}-A_{R,\mu}^{a}) (A_{L,\nu}^{b}-A_{R,\nu}^{b})
%\end{equation}
%The metric then takes the form:
\begin{equation}
	ds^2_{\text{Id}/\alpha}=\frac{l^2}{r^2}dr^2 \pm \mu\nu  \left(r^2+\frac{1}{r^2}\frac{L(x^+)}{\mu^2 }\frac{\tilde{L}(x^-)}{\nu^2 }\right)dx^+ dx^- +L(x^+) (dx^+)^2+\tilde{L}(x^-) (d x^-)^2
 \label{MetricSolutions}
\end{equation}
% \begin{equation}
% 	ds^2=\frac{dr^2}{r^2}+\frac{M}{2}\left(r^2+\frac{4}{M^2}\frac{L(x^+)\tilde{L}(x^-)}{r^2}\right)dx^+ dx^- +L(x^+) (dx^+)^2+\tilde{L}(x^-)^2 (d x^-)^2
% \end{equation}
where the tensor components $L$ and $\tilde{L}$ are :
\begin{align}
	L=-\frac{1}{2}\{F,x^+\} \, , \qquad 
	\tilde{L}=-\frac{1}{2}\{\tilde{F},x^-\} \, .
 %=\frac{3(\tilde{F}'')^2-2\tilde{F}''' \tilde{F}'}{4(\tilde{F}')^2}
\end{align}
The upper sign in equation (\ref{MetricSolutions}) corresponds to the choice of identity automorphism while the lower sign corresponds to the choice of outer automorphism $\alpha$. Only the sign of $\mu\nu$ plays a role in the metric -- its absolute value can always be absorbed in the coordinates $x^+,x^-$ and 
the functions $L, \tilde{L}$, or shifted freely by a Liouville field redefinition.
Note that even the leading asymptotic behaviour and therefore the boundary metric itself depends on the choice of outer automorphism $\alpha$. 
The functions $L$ and $\Tilde{L}$ appearing in the metric are the components of the stress-energy tensor of Liouville 
theory. Indeed, plugging the solutions (\ref{LiouvilleSolutions}) in the stress-energy tensor, one finds:
\begin{equation}
    T_{++}=-\frac{1}{2b^2}\{F,x^+\}=\frac{L}{b^2} 
    \, , \qquad  T_{--}=\frac{\Tilde{L}}{b^2} \, .
\end{equation} 
The fact that fluctuations around
the AdS metric can be parameterized in terms of a (e.g. Liouville) stress-energy tensor is well-known  \cite{Skenderis:1999nb}.

\subsection{The Stability and the Metric}
We combine the facts we reviewed about the classical relation between Liouville theory, Wess-Zumino-Witten models, Chern-Simons theories and Einstein gravity with our observation in section \ref{OuterAutomorphism} on the ambiguity in the definition of the Chern-Simons formulation of gravity. This allows us to make a crucial observation. 
If we wish our asymptotic metric (\ref{MetricSolutions}) to take a standard $AdS_3$ form in which the boundary time is identified with the Liouville time $t$, then the product $\mu \nu=b^2 M$ needs to be negative if we choose our automorphism to be the identity and it needs to be positive for the choice of a non-trivial outer automorphism $\alpha$.
In other words, in order for the classical Liouville theory to be stable and the metric theory to be asymptotically $AdS_3$ with a standard induced boundary time, we must choose the Chern-Simons formulation of gravity with the non-trivial outer automorphism identification.
This is a crucial  observation. 

Let us briefly discuss proposals that would render the theory with the identity automorphism internally consistent. An existing  proposal  is to perform a non-local field redefinition in Liouville theory in order to flip the sign of the potential \cite{Henneaux:1999ib}. Unfortunately, even if one could track the field redefinition in the quantum path integral, this is  bound to interfere with the unitarity of the boundary theory since the operation relates a potential which is bounded from below to one that is unbounded from below.
Another proposal is to perform an analytic continuation in the parameter $M$. Correlators in Liouville theory have a very mild dependence on this parameter and one may be optimistic that one can implement such an analytic continuation, but again, one would need to continue the unitary structure of the theory as well. Finally, one can act on the gravitational side of the analysis and propose an analytic continuation of all metrics in the radial coordinate, $r \rightarrow ir$. That too is an interesting option. 

In this paper, we will stick to identifying the Chern-Simons and gravity formulations using the non-trivial outer automorphism $\alpha$, guaranteeing 
the stability and the unitarity of the boundary Liouville theory.

\subsection{The Classical Solutions with Hyperbolic Monodromy}
\label{ClassicalGravitySolutions}
Our boundary Liouville conformal field theory is equipped with a unitary Hilbert space with fixed  spectrum \cite{Teschner:2001rv}. The Hilbert space corresponds to the quantization of a well-defined subset of classical solutions. Since we wish to define our bulk theory of gravity in terms of the boundary Liouville conformal field theory, we will accept that our bulk theory of gravity is a quantization of a definite subset of bulk metric solutions. It is certainly interesting in the context of attempts to quantize gravity to explicitly state which subsets of metrics this is. 
The solutions that give rise to the quantum Hilbert space of Liouville theory are those
with hyperbolic monodromies. Thus, we follow these solutions along our equivalence maps and state which on-shell metrics correspond to Liouville solutions with these monodromies. 

The primary hyperbolic solutions for a space-time with cylindrical boundary are given in terms of the chiral functions
-- see e.g. \cite{Seiberg:1990eb} --: 
		% \begin{align}
		% 	F(x^{+})=(x^{+})^{a}\\
		% 	F(x^-)=(x^-)^{a}
		% \end{align}
  % or  decide on boundary space 
  \begin{equation}
F = \exp( a x^+)
\, , \qquad 
\tilde{F} = \exp( a x^-) \, .
  \end{equation}
  To obtain a periodic Liouville field on the cylinder, it is crucial that the left and right monodromy $a$ be equal. 
		These correspond to tensors:
		% \begin{align}
		% L_+=\frac{a^2-1}{4 (x^{+})^2}, \ L_-=\frac{a^2-1}{4 (x^-)^2}
		% \end{align}
  %or 
  \begin{align}
		L=\frac{a^2}{4 } \, , \qquad \tilde{L}=\frac{a^2}{4 } \, .
		\end{align}
  For completeness, we review how these metrics are related to the BTZ black holes \cite{Banados:1992gq}. In Fefferman-Graham coordinates \cite{FG}, the BTZ metric reads:
% \begin{equation}
% ds^2 = \frac{d \rho^2}{\rho^2} + \rho^2 (dw^+ + \frac{(r_++r_-)^2}{\rho^2} dw^-) (dw^- + \frac{(r_+-r_-)^2}{\rho^2} dw^+)
% \end{equation}
% where $w^\pm = (\phi \pm t)/2$.
% In our boundary coordinates then:
\begin{equation}
ds^2 = \frac{d \rho^2}{\rho^2} - \frac{\rho^2}{4}  (dx^+ -\frac{(r_+-r_-)^2}{\rho^2} dx^-) (dx^- - \frac{(r_++r_-)^2}{\rho^2} dx^+) \, .
\end{equation}
This is identical to the metric solution (\ref{MetricSolutions}) with outer automorphism $\alpha$ when we identify the parameters as:
\begin{align}
L &= \frac{(r_+ + r_-)^2}{4} =\frac{a^2}{4}
\nonumber \\
\tilde{L} &= \frac{(r_+ - r_-)^2}{4} = \frac{a^2}{4} \, .
\end{align}
Thus, the primary hyperbolic solutions have zero spin \cite{Li:2019mwb},
as do the primaries in the Liouville theory spectrum, and they correspond to BTZ metric solutions with mass
\begin{equation}
M = r_+^2
%+r_-^2 \, , \qquad J=2 r_+ r_- \, .
%\end{equation}
%So, or $J=0$, we have
%\begin{equation}
%M 
= a^2 \, .
\end{equation}
The energy-momentum tensor of Liouville theory as well as the energy-momentum tensor in the gravity solution equals 
\begin{equation}
T_{++} = \frac{a^2}{4 b^2}
%= \frac{L}{b^2} 
\, 
\end{equation}
in the classical limit. 
%\subsection{The Boundary Gravitons}

We can moreover use the B\"acklund transformation from Liouville theory to a free field theory to parameterize the Liouville energy-momentum tensor in terms of a free field energy-momentum tensor (supplemented with a linear dilaton). The hyperbolic monodromy becomes the momentum of the free field while the boundary gravitons correspond to Fourier modes excited in the free field energy-momentum tensor.\footnote{Details of this fact are discussed in the periodic case i.e. of a cylindrical boundary in e.g. \cite{Braaten:1982fr,Teschner:1995dt}.}
Thus, we know exactly which space of classical Liouville solutions corresponds to the quantum Hilbert space and are able to translate this space into a set of classical solutions of three-dimensional gravity: they are the BTZ black holes with positive mass dressed with boundary gravitons \cite{Li:2019mwb}.

\section{Liouville Three-dimensional Quantum  Gravity}
\label{QuantumLiouvilleGravity}
In this subsection we review and complete the evidence that the classical equivalences we tracked can be promoted to dualities of the quantum theories. Most of the individual  links between theories are known \cite{Coussaert:1995zp}, but the detailed chain of reasoning that leads to an equivalence between three-dimensional gravity and Liouville theory in the quantum theory deserves a review. 

\subsection{From Liouville to Chiral Wess-Zumino-Witten Models}
Firstly, we discuss the quantum incarnation of the first step in our equivalence path drawn in Figure \ref{Path}. A generic reasoning at each step is to state that the classical solution spaces that we identified (consisting of primary hyperbolic monodromies and descendants, or equivalently, of black holes dressed with boundary gravitons) are identical, that the symplectic forms coded in their equal actions are the same and that therefore we can choose an equivalent quantization for all of these phase spaces. This is a solid reasoning to identify the Hilbert spaces and through the action, the interactions as well. Our discussion of the individual steps  fleshes out this sturdy skeleton. 

The path integral equivalence of the Liouville and the null gauged Wess-Zumino-Witten model 
\begin{equation}
\int dg dC_L dC_R e^{iS_{GWZW}[g,C_L,C_R]}  = \int d \phi e^{iS_{\text{Liouville}}[\phi]} 
\end{equation}
was discussed in great detail in 
\cite{ORaifeartaigh:1997wnu,ORaifeartaigh:1998pox}.  Keeping careful track of the measures in the path integration indeed gives rise to the quantum mechanically exact identification 
\begin{equation}
\sqrt{k-2}=b^{-1} \, 
\end{equation}
of the respective coupling constants in the actions.
It is possible to further improve on this path integral calculation of the partition function \cite{ORaifeartaigh:1997wnu,ORaifeartaigh:1998pox} by performing the computation in the presence of gauge invariant insertions. For this analysis, it may well be useful to exploit the map between the Euclidean $H_3^+$  and Liouville conformal field theory correlators 
\cite{Ribault:2005wp},
in particular in its path integral derivation \cite{Hikida:2007tq} as a starting point for the null gauging. 
The correspondence involves the standard  map between $\mathfrak{sl}(2,\mathbb{R})$ spins $j=1/2+is$ and Liouville momenta $\alpha=Q/2+i p$:
\begin{equation}
p = s/\sqrt{k-2} \, .
\end{equation}

The second step which involves the splitting of the (gauged) Wess-Zumino-Witten model into two chiralities is also discussed in great detail in the literature in the path integral formalism in
\cite{Stone:1989vg} and includes a careful discussion of the gauge symmetry in \cite{Harada:1990tw}. Indeed, these references show that the classical treatment in our Appendix \ref{Split} can be carried over to the quantum path integral. The basic but central observation is that the following path integrals match:
\begin{equation}
\int \frac{dg_L dg_R}{\text{Gauge Group Volume}}
e^{iS_{WZW}[g_L g_R]} = \int dg e^{iS_{WZW}[g]}
\,  \, .
\end{equation}
The only small difference in our quantum equivalence between chiral and non-chiral models  is that we have a null gauge symmetry, but the adaptation is entirely straightforward. To claim this though, our off-shell treatment of the splitting procedure in Appendix \ref{Split} is paramount, and this is one crucial distinction between our analysis and those performed previously. It is important to note that the gluing gauge symmetry ensures that the spin $j$ in the gauged Wess-Zumino-Witten model is diagonal, as is the Liouville momentum $p$ in the Liouville path integral.\footnote{We already saw from the classical Liouville solution on the cylinder that the periodic nature of the field enforces equal left and right monodromy.}

Moreover, the (Lagrangian) path integral splitting is  complemented by a Hamiltonian Drinfeld-Sokolov reduction procedure that leads to the same quantum equivalence of theories. The latter manifestly matches the quantum Virasoro symmetries of the models \cite{Bershadsky:1989mf} and the equivalence then boils down to a matching of the spectrum and correlation functions which has been analyzed in detail in the literature \cite{Bershadsky:1989mf,Ribault:2005wp,Hikida:2007tq}. Various subtleties in these equivalences (like the insertion of degenerate operators in correlation functions) may deserve further scrutiny in the gravitational setting.
It should be noted that our equivalences are mild extensions of those in the literature. E.g. the term linear in the boundary gauge fields $C_{L,R}$ in the chiral/anti-chiral Wess-Zumino-Witten models has to be carried along in the analysis. The associated path integrations are algebraic and have trivial measure factors associated to them and exact quantum constraints -- they hardly complicate the proofs of equivalence.

Up to this stage, the equivalences are between two-dimensional conformal field theories. They are rather rigorously established for instance in terms of operator realizations of the left and right Virasoro symmetry algebras (e.g. in terms of coset conformal field theories). An illustration is the equivalence of the quantum energy-momentum tensors
\begin{equation}
T_{\text{Liouville}} = T_{\mathfrak{sl}(2,\mathbb{R})} + T_{\text{ghosts}} + T_{\text{improv}} \, ,
\end{equation}
where the formula is a schematic representation of the Liouville, the (chiral) Wess-Zumino-Witten model, the (gauge fixing) ghosts and the improvement energy momentum tensors -- see e.g. \cite{Bershadsky:1989mf} for all the details. Rather, the non-trivial part of the equivalence is to appropriately choose the Hilbert space in each guise of the theory -- we do this by fiat. 

\subsection{From  Wess-Zumino-Witten Models to Chern-Simons Theories}

The equivalence of the chiral boundary Wess-Zumino-Witten model with a Chern-Simons theory was established at the quantum level in \cite{Witten:1988hf,Elitzur:1989nr}.
The extension of the quantum equivalence to the bulk involves the field redefinition from the three-dimensional gauge field components $\tilde{A}_L$ to the group valued field $g_L$:
\begin{equation}
\int D \tilde{A}_L \delta (\tilde{F}_L) e^{iS_{CS}[\tilde{A}_L]} \dots = \int dg_L  e^{S_{cWZW}[g_L]} \dots
\end{equation}
where we made explicit the constraint on the field strength components $\tilde{F}_L$ solved for e.g. in equation (\ref{SolutionConstraint}).
This field redefinition has unit Jacobian \cite{Elitzur:1989nr} and therefore the quantum measure on both sides of the equivalence is standard. To this  analysis we  add the boundary gauge field term which goes along for the ride once more. 
To reach the Chern-Simons theories with polar boundary conditions, we performed a gauge transformation which is a symmetry in the quantum theory. 

Thus, we wind up with a quantum equivalence between Liouville theory and Chern-Simons theories which have prescribed radial boundary conditions.
As stressed in the treatment of the classical theory, we still have a gauged boundary symmetry which serves to glue the left and right  modes in the chiral and anti-chiral theories.

\subsection{From Chern-Simons to Metric Gravity}
As far as quantum equivalences are concerned, only the equivalence between the integration over Chern-Simons gauge fields and metrics is hard to fathom. This is a main point of our approach: the quantum theory of gravity is defined to correspond to the standard integration measure in Liouville theory \cite{Li:2019mwb}. We can be more precise about what this entails. In our final step, we start out with an integration over gauge fields $A_{L,R}$ with standard measure $dA_{L,R}$ and chiral boundary conditions supplemented with those imposed by a functional integration over the boundary gauge fields $C_{L,R}$. 
The change of variables from the connections $A_{L,R}$ to the dreibein $e$ and spin connection $\omega$ is linear and has trivial Jacobian.
\begin{equation}
\int dA_L d A_R e^{iS_{CS}[A_L]-iS_{CS}[A_R]} \dots
= \int d\omega de e^{iS_{EH}[e,\omega]} \dots
\end{equation}
Classically, the zero torsion and zero curvature equations match the zero Chern-Simons curvature equations of motion. The left hand side  arose from the Liouville path integral with standard Liouville measure. The right hand  gravitational path integral also has a natural measure, in the Palatini formulation of gravity.  The metric formulation of this integration measure is considerably more complicated. 
It was already remarked in \cite{Witten:1988hc} that the integral over the Chern-Simons fields or the spin connection and dreibein includes an integral over degenerate metrics. It was proposed that metric gravity can be thought off as a broken phase of Chern-Simons gravity. We note that in asymptotically $AdS_3$ space-times, the situation is intermediate in the following sense. The boundary values of the metric are fixed and non-degenerate while it remains true that in the interior of the space-time, we are also integrating over possibly degenerate metric fluctuations.  

Our prescription is considerably more specific though. We define a theory in which we integrate over metric fluctuations determined by the Liouville field through the energy-momentum operator with a measure determined by the Liouville field itself (and not as a naive approach might suggest, by the metric fluctuation). 
Moreover, as stressed on several occasions, we give a clear prescription on which metrics to allow in the integration and which not. 
Indeed, we want to stress that our theory allows for singular metrics in the same sense that Liouville theory allows for the integration over singular configurations once vertex operators are inserted. On the bulk side, if one allows for black hole metrics (as we do), one allows for the same type of singularities in the path integration. 
In the Liouville path integral, on the boundary torus for instance, one integrates over well-defined (for instance doubly periodic) configurations. These include configurations that can be regularly extended to gauge fields in the bulk as well as configurations that have monodromies around non-trivial boundary cycles that are contractible in the bulk. The latter are singular bulk configurations -- by definition we include them in our integration measure. Thus, we can intuitively describe the $(A_L,A_R)$ or $(e,\omega)$ configurations over which we integrate as having
a diagonal hyperbolic monodromy and otherwise entirely regular in the bulk. This is true classically, as in subsection \ref{ClassicalGravitySolutions} as well as quantum mechanically \cite{Li:2019mwb}. 

We can obtain an implicit off-shell description of the metric (and torsion) we are integrating over, as follows. 
We  explicitly solve the constraint:
\begin{equation}
(\partial_x g_L g_L^{-1})^- = \mu
\end{equation}
and gauge fix the diagonal components of $A_{L,x}=-\partial_x g_L g_L^{-1}$ to zero. We then find that the left gauge field is given in terms of the group valued field $g_L$:
\begin{equation}
g_L=\sqrt{\frac{\mu}{\partial_x y_L}}
    \begin{pmatrix}
	   \frac{y_L \partial_x^2 y_L-2 (\partial_x y_L)^2 }{2\mu \partial_x y_L} &-\frac{\partial_x^2 y_L}{2\mu \partial_x y_L} \\
		y_L & 1
	\end{pmatrix},
\end{equation}
where $y_L(t,x)$ is the coefficient of $t_-$ in the Gauss decomposition of $g_L$. A similar equation holds for the right-movers. We can then parameterize the metric in terms of the chiral Liouville fields off-shell:
\begin{multline}
	ds^2=\frac{dr^2}{r^2} \pm \mu \nu\left(r^2+\frac{b^4}{\mu^2 \nu^2 r^2}\left(\frac{1}{2}(\partial_x \phi_L)^2-\frac{1}{\sqrt{2} b}\partial_x^2 \phi_L\right)\left(\frac{1}{2}(\partial_x \phi_R)^2-\frac{1}{\sqrt{2}b}\partial_x^2 \phi_R \right)\right)dx^+ dx^- \\ 
 +b^4\left(\frac{1}{2}(\partial_x \phi_L)^2-\frac{1}{\sqrt{2} b}\partial_x^2 \phi_L \right)^2 (dx^+)^2+b^4\left(\frac{1}{2}(\partial_x \phi_R)^2-\frac{1}{\sqrt{2} b}\partial_x^2 \phi_R \right)^2 (d x^-)^2
\end{multline}
Possible terms arising from the gauge field components $A_{L,-}$ and $A_{R,+}$ only start to contribute at order $1/r^4$.
There is an implicit link between the fields $\phi_L$ and $\phi_R$ and the Liouville field $\phi$ via the set of equations:
\begin{align}
e^{{\sqrt{2}} {b}\phi} &= \frac{1}{b^2 M}\frac{\partial_x y_L \partial_x x_R}{1+x_R y_L} \nonumber \\
e^{{\sqrt{2}} {b}\phi_L} &= \frac{1}{\mu}\partial_x y_L \nonumber \\
e^{{\sqrt{2}} {b} \phi_R} &= \frac{1}{\nu}\partial_x x_R  \, .
\end{align}
One can check that this metric reduces to our earlier expression (\ref{MetricSolutions}) on-shell . Our final specification 
is that it is the field $\phi$ which is fundamental and well-defined on the boundary surface.

We can %
point to several properties of the theory we define which differ from other metric quantizations of pure gravity. 
One difference is that in the quantum Chern-Simons theory, we path integrate over configurations with torsion. Indeed, this is rather intrinsic to the Chern-Simons formulation of quantum gravity. 
We note that on-shell  configurations do have zero torsion. Even when they have monodromy, they have equal left and right hyperbolic monodromy and still have zero torsion. 
 Moreover, these on-shell configurations represent the quantum Hilbert space well. 
A second observation is that the Liouville field parameterizes a one-dimensional subspace of multiple Einstein equations. In particular, when the Liouville stress tensor is off-shell, it can  be non-conserved. These are two classical equations which are quantized along a one-dimensional slice parameterized by the Liouville field. 
Finally, note that the quantization is over metrics that factorize into a left- and a right-moving part such that the Virasoro algebras are neatly separated. Modular invariance is guaranteed by gluing. 

\section{Conclusions}
\label{Conclusions}
In this paper, we further analyzed the link between the Liouville conformal field theory on the boundary and a theory of pure gravity with negative cosmological constant in the bulk \cite{Coussaert:1995zp}. We demonstrated that the classical arguments that relate the action principles need to be twisted by an outer automorphism of the $\mathfrak{sl}(2,\mathbb{R})$ algebra in order to make two requirements consistent: that the boundary theory be stable and that the metric of the boundary theory agree with the metric induced from the bulk. In this manner, we constructed a stable Lorentzian Liouville theory dual to a well-defined set of  metrics in the bulk.  
We moreover proved the equivalence of the theories off-shell which is important to lift it to the quantum theory. Our detailed derivation
renders the proposal of the quantum equivalence of the theories and its consequences \cite{Li:2019mwb} even more concrete. The unitary Liouville conformal field theory has a well-defined spectrum that can be traced back to a particular subset of classical solutions. It is these solutions that we tracked to the gravitational bulk in order to argue clearly which set of metrics is quantized in the dual Liouville theory. This confirms the salient features of the quantum Liouville theory of pure three-dimensional gravity discussed in \cite{Li:2019mwb} and renders it stable. 

We added comments on the boundary conditions of asymptotic $AdS_3$ space-times in the Chern-Simons formulation. Firstly, we argued that an infrared regularization of space-time is necessary in order to relate those boundary conditions to the standard null gauging of the Wess-Zumino-Witten model. Secondly, we established a partial analogy with oper boundary conditions and stressed the relation between the single pole gauge field boundary conditions and the double pole metric boundary conditions in asymptotically $AdS$ space-times.
We also emphasized the importance of the gauging of a left-right multiplication symmetry in order to glue the chiral-anti-chiral description of the theory.

There are many open directions for further research. It  would be interesting to perform an analogous classical and quantum analysis in the Euclidean. Since the local gauge group is different, there are many important details that will change. Note that in the Euclidean setting, 
 there is no standard Euclidean Weyl factor instability \cite{Gibbons:1978ac} due to the topological bulk on the one hand, and the fact that the leading order asymptotic metric is fixed. Thus, one may have an entirely well-defined Euclidean path integration problem. 
 It would also be worthwhile to match the semi-classical analysis of Liouville correlators to a semi-classical analysis of the behaviour of classical solutions in three-dimensional gravity, in particular in regard to black hole merger. 
 Moreover, one can extend our analysis to more general leading order boundary metrics. We also note that
 when the boundary metric is curved, the inclusion of the outer automorphism again ensures an accord between the induced metric on the boundary and the metric used to define the Liouville action. This follows from combining our analysis with the results of e.g. \cite{Rooman:2000zi}.

Finally, let us recall a few salient features of our three-dimensional theory of pure quantum gravity with negative cosmological constant \cite{Li:2019mwb}. 
We differ in the ground state with most of the literature in that we consider the  conformal invariant vacuum to be non-normalizable -- the conformal field theory spectrum is gapped. Moreover, our spectrum is continuous and consists solely of spinless (primary) black holes, dressed with (descendant) boundary gravitons. Due to the topological nature of pure three-dimensional gravity there is no Hawking radiation which may be related to a lack of a statistical mechanical interpretation of a tentative thermodynamics. Thus, this theory is isolated from e.g. the string theories in $AdS_3$ that have largely the opposite properties. An interesting exception is string theory in $AdS_3$ at high curvature where the $SL(2,\mathbb{C})$ invariant ground state becomes non-renormalizable \cite{Giveon:2005mi}. 

It would certainly be interesting to extend our interpretation of this seemingly isolated theory of quantum gravity to other set-ups. An example is to extend it to spaces with  two asymptotic regions \cite{Henneaux:2019sjx}. In as far as the boundary gravitons only are concerned, the recent contribution
\cite{Collier:2023fwi} computes their partition function in greater generality. It would be interesting to add the spectrum of black holes states to these analyses and further explore the properties of pure quantum gravity with negative cosmological constant on more general space-times with self-consistent conformal field theory duals.

\appendix

\section{Conventions}
\label{Conventions}
The $\mathfrak{sl}(2,\mathbb{R})$ Lie algebra has signature $(-,+,+)$ in the Killing metric. It has the commutation relations:
\begin{align}
[ t_2, t_{\pm}] &= \pm  t_{\pm} \, , \qquad \qquad 
[t_+, t_-] =  2 t_2 \, .
\end{align}
We denote the generators by $t_a$.
% and the structure constants:
% \begin{align}
% [t_a, t_b] &= {f_{ab}}^c t_c
% \, .
% \end{align}
A two-dimensional representation is given by
\begin{equation}
	t_{+}=
    \begin{pmatrix}
	0&1\\
	0&0
	\end{pmatrix}, \quad
	t_{-}= 
    \begin{pmatrix}
	0&0\\
	1&0
	\end{pmatrix}, \quad
	t_2= \frac{1}{2}
    \begin{pmatrix}
	1&0\\
	0&-1
	\end{pmatrix} \, .
\end{equation}
We have that:
\begin{equation}
\Tr (t_2 t_2)=\frac{1}{2} \, , 
\qquad 
\Tr (t_+ t_-)= 1
\end{equation}
and the convention that the non-zero components of the metric in the Lie algebra are:
\begin{equation}
\eta_{22}=1 \, , \qquad \eta_{+-}=2 \, 
\end{equation}
such that the formula
\begin{equation}
\Tr (t_a t_b) = \frac{1}{2} \eta_{ab}  \, 
\end{equation}
holds.

\section{Splitting a Gauged Wess-Zumino-Witten Model}
\label{WZWSplit}
\label{Split}
This Appendix demonstrates how to separate a null gauged Wess-Zumino-Witten model into a chiral and an anti-chiral gauged Wess-Zumino-Witten model. The Appendix makes reasonings spread broadly in the literature more readily accessible. 

\subsection{The Gauged Wess-Zumino-Witten Model}
The first matter to recall is that when one gauges a Wess-Zumino-Witten model classically, it is a challenge to render the Wess-Zumino term gauge invariant. In fact, this can only be achieved in particular circumstances, amply discussed in the Appendix to \cite{Witten:1991mm}. We are in the circumstance in which both the single left generator and the single right generator we gauge are null. In that favorable case, the naive gauging of the Wess-Zumino-Witten model goes through, without the need for a local counterterm, and the gauged action is straightforwardly obtained  \cite{Balog:1990mu}.

Firstly, one records the Polyakov-Wiegmann identity, which in our conventions reads:
\begin{equation}
S_{WZW}[g_L g_R]=S_{WZW}[g_L] + S_{WZW}[g_R] +\frac{k}{\pi}\int d^2 x\ \Tr[g_L^{-1} \partial_- g_L \partial_+ g_R g_R^{-1} ] \, .
\end{equation}
The cross-term arises from a (symmetric) kinetic term contribution and an (anti-symmetric) Wess-Zumino Stokes term. These combine with equal coefficients to give a term consistent with the chiral global symmetry of the Wess-Zumino-Witten model.  
Secondly, for the gauging of the null left and right symmetries, one proposes the action:
\begin{align}
S_{GWZW}(g,h,\tilde{h}) &= S_{WZW}(h g \tilde{h}^{-1})
\end{align}
where the left null group element $h$ and the right null group element $\tilde{h}$ are seen as parameterizations of the corresponding (chiral) gauge fields $C_{L,R}$. The action has a manifest gauge symmetry by left and right multiplication of the group valued map $g$ combined with right and left multiplication of $h$ and $\tilde{h}$. One computes the action through application of the Polyakov-Wiegmann identity and finds:
%Apparently, various spellings exist for paramet(e)rization.
\begin{align}
S_{GWZW}(g,h,\tilde{h})
%&= S_{WZW}(hg) + \frac{k}{\pi} \int d^2 x \Tr[ g^{-1} h^{-1} \partial_- (hg) \partial_+ \tilde{h}^{-1} \tilde{h} ]
%\nonumber \\
&= S_{WZW}(g)  \\
& + \frac{k}{\pi} \int d^2 x \Tr[
h^{-1} \partial_- h \partial_+ g g^{-1} -
g^{-1} h^{-1} \partial_- h g   \tilde{h}^{-1} \partial_+ \tilde{h}
- g^{-1} \partial_- g  \tilde{h}^{-1} \partial_+ \tilde{h}
] \, . \nonumber
\end{align}
The null Wess-Zumino-Witten actions $S_{WZW}(h)$ and $S_{WZW}(\tilde{h})$ are zero.  We now introduce the gauge field components:
\begin{align}
C_L t_+ &=h^{-1} \partial_- h
\nonumber \\
C_R t_- &=  \tilde{h}^{-1} \partial_+ \tilde{h}  \, ,
\end{align}
and find the action  \cite{Balog:1990mu,Forgacs:1989ac}:
\begin{align}
S_{GWZW}(g,h,\tilde{h}) &=
S_{WZW}(g) + \frac{k}{\pi} \int d^2 x \Tr[
C_L t_+ \partial_+ g g^{-1}
- C_R t_- g^{-1} \partial_- g 
- g^{-1} C_L t_+ g C_R t_- 
] \, . \nonumber
\end{align}
This agrees with the action in the bulk of the paper, up to the extra terms proportional to the constants $(\mu,\nu)$ which can be added by hand -- they are separately gauge invariant for the abelian gauge algebra. 
\subsection{The Split}
In this subsection, we wish to demonstrate the equivalence between the chiral/anti-chiral and the ordinary gauged Wess-Zumino-Witten model. We follow the technique described in \cite{Stone:1989vg,Harada:1990tw}.
We split the field $g$ in the gauged action $S_{GWZW}$ into two further group elements $g_L$ and $g_R$ and introduce a gauge field component $B$
\cite{Stone:1989vg,Harada:1990tw}:
\begin{align}
S_{cGWZW}(hg_L,g_R \tilde{h}^{-1};B) &= S_{WZW} (h g_L g_R \tilde{h}^{-1}) 
\nonumber \\
& - \frac{2k}{ \pi} \int d^2 x \Tr (B-\frac{1}{2} (\partial_+ (g_R \tilde{h}^{-1} ) (g_R\tilde{h}^{-1})^{-1} + (hg_L)^{-1} \partial_- (hg_L)) )^2 \, .
\end{align}
It is clear that if we consider $B$-independent observables and perform the quadratic integration over $B$, then we obtain a model with a gauge invariance $(g_L,g_R) \rightarrow (g_L k^{-1},k g_R)$ which is equivalent to the original gauged model $S_{GWZW}$. Alternatively, one can choose the gauge $B=0$ which gives rise to a field-independent determinant factor that one can neglect \cite{Stone:1989vg,Harada:1990tw}. In this gauge, we find that the model is equivalent to two chiral gauged Wess-Zumino-Witten models.
The point of the square we add is to kill a Polyakov-Wiegmann cross term and to render the two other terms chiral.  In practice, this goes through
the  calculation:
\begin{align}
S_{cGWZW}(hg_L,g_R \tilde{h};0) &= S_{WZW} (h g_L g_R \tilde{h}^{-1}) \nonumber \\
& - \frac{k}{2 \pi} \int d^2 x \Tr  (\partial_+ (g_R \tilde{h}^{-1} ) (g_R\tilde{h}^{-1})^{-1} + (hg_L)^{-1} \partial_- (hg_L))^2)  \nonumber \\
&= S_{WZW}(h g_L) + S_{WZW}(g_R \tilde{h}^{-1}) \nonumber \\
&
- \frac{k}{2 \pi} \int d^2 x \Tr  ( \partial_+ (g_R \tilde{h}^{-1}) (g_R\tilde{h}^{-1})^{-1} )^2 + \left( (hg_L)^{-1} \partial_- (hg_L))^2 \right) 
\nonumber \\
&= S_L(h g_L) +S_R(g_R \tilde{h}^{-1}) 
\, ,
\end{align}
where we defined the chiral actions:
\begin{equation}
S_L(g_L) = S_{WZW}(g_L) - \frac{k}{2 \pi} \int d^2 x \Tr (g_L^{-1} \partial_- g_L g_L^{-1} \partial_- g_L) 
\end{equation}
and
\begin{equation}
S_R(g_R) = S_{WZW}(g_R) - \frac{k}{2 \pi} \int d^2 x \Tr (g_R^{-1} \partial_+ g_R g_R^{-1} \partial_+ g_R)  \, .
\end{equation}
Again, the null group fields $h$ and $\tilde{h}$ are to be identified with gauge field components. Indeed, the gauged chiral left action is:
\begin{align}
S_L(h g_L) &= S_{WZW}(hg_L) -   \frac{k}{2 \pi} \int d^2 x \Tr ((hg_L)^{-1} \partial_- (h g_L) (h g_L^{-1}) \partial_- (h g_L))
\nonumber \\
&= S_{WZW}(h) + S_{WZW}(g_L) + \frac{k}{\pi} \int d^2 x \Tr[ h^{-1} \partial_- h \partial_+ g_L g_L^{-1}]
\nonumber \\
& -  \frac{k}{2 \pi} \int d^2 x \Tr (g_L^{-1} \partial_- g_L + g_L^{-1} h^{-1} \partial_- h g_L)^2
\nonumber \\
% &= S_L(g_L) + \frac{k}{\pi} \int d^2 x \Tr[ h^{-1} \partial_- h (\partial_+-\partial_-) g_L g_L^{-1}]
% + 0
% \nonumber \\
&= S_L(g_L) + \frac{k}{\pi} \int d^2 x \Tr[ C_L t_+ \partial_x g_L g_L^{-1} ] \nonumber \\
&= \frac{k}{2 \pi} \int d^2 x \Tr [ 
g_L^{-1} \partial_- g_L g_L^{-1} \partial_x g_L
]+S_{WZ}(g_L)
\nonumber \\
& + \frac{k}{\pi} \int d^2 x \Tr[ C_L t_+ \partial_x g_L g_L^{-1} ]\, .
\end{align}
For the right-movers, we similarly find for the anti-chiral gauged action:
\begin{align}
S_R (g_R \tilde{h}^{-1})
&= S_{WZW}(g_R \tilde{h}^{-1})- \frac{k}{2 \pi} \int d^2 x \Tr [ ((g_R \tilde{h}^{-1})^{-1} \partial_+(g_R \tilde{h}^{-1}))^2  ]
\nonumber \\
&= S_{WZW}(g_R) + \frac{k}{\pi} \int d^2 x \Tr [ g_R^{-1} \partial_- g_R \partial_+ (\tilde{h}^{-1}) \tilde{h}]
\nonumber \\
& - \frac{k}{2 \pi} \int d^2 x \Tr ( -\partial_+ \tilde{h} \tilde{h}^{-1} + \tilde{h} g_R^{-1} \partial_+ g_R \tilde{h}^{-1}   )^2
\nonumber \\
&= S_R(g_R) + \frac{k}{\pi} \int d^2 x \Tr[ C_R t_- g_R^{-1} \partial_x g_R ]
\nonumber \\
&= -\frac{k}{2 \pi} \int d^2 x \Tr [ g_R^{-1} \partial_+ g_R g_R^{-1} \partial_x g_R  ] +S_{WZ}(g_R) \nonumber \\
& + \frac{k}{\pi} \int d^2 x \Tr[ C_R t_- g_R^{-1} \partial_x g_R ] \, .
\end{align}
Again, the terms proportional to the constants $(\mu,\nu)$
can be added to the action. 

\bibliographystyle{JHEP}

\end{document}